\documentclass[12pt,preprint]{aastex}

\usepackage{emulateapj5}
\usepackage{amssymb}
\usepackage{amsmath}
\usepackage{amsfonts}
\usepackage{graphicx}

\newcommand{\hscale}{\ensuremath{\,h_{70}^{-1}}}  % Symbol for Hubble 
\newcommand{\lya}{\ensuremath{\rm Ly\alpha\;}}   % Symbol for Lyman-alpha
\newcommand{\kms}{\ensuremath{\rm \,km\,s^{-1}}} % Symbol for km/s
\newcommand{\kpc}{\ensuremath{\rm \,kpc}} % Symbol for kpc
\newcommand{\cmsq}{\ensuremath{\rm \,cm^{-2}\;}}   % Symbol for cm^-2
\newcommand{\hnot}{\ensuremath{ H_{0}}}       % Symbol for Hubble constant
\slugcomment{Draft Version \today}
\begin{document}

\title{The Local \lya Forest: Absorbers in Galaxy Voids}

\author{Kevin M. McLin\altaffilmark{1,2,3}, John T. 
Stocke\altaffilmark{1,2,3}, R. J. Weymann\altaffilmark{4},
Steven V. Penton\altaffilmark{1,2}, \mbox{J. Michael Shull\altaffilmark{1,5}}}

\email{mclin@casa.colorado.edu, stocke@casa.colorado.edu, rjw@ociw.edu, 
spenton@casa.colorado.edu, mshull@casa.colorado.edu}

\altaffiltext{1}{Center for Astrophysics and Space Astronomy and Dept. of
Astrophysical and Planetary Sciences, University of 
Colorado, Boulder, CO 80309}
\altaffiltext{2}{Visiting Astronomer, Kitt Peak National Observatory, National 
Optical Astronomy Observatories, which is operated
      by the Association of Universities for Research in Astronomy, Inc. 
(AURA) under cooperative agreement with the
      National Science Foundation.}
\altaffiltext{3}{Visiting Astronomer, Las Campanas Observatory}
\altaffiltext{4}{Carnegie Observatories, 813 Santa Barbara St., Pasadena, CA 
91101}
\altaffiltext{5}{also at JILA, University of Colorado and National Institute 
of Standards and Technology, Boulder, CO 80309}

\begin{abstract}

We have conducted pointed redshift surveys for galaxies in the direction of bright AGN whose HST far-UV spectra contain nearby (${ cz
\lesssim 30,000\kms}$), low column density (${ 12.5 \leqslant \log\, N_{HI} (\cmsq) \leqslant 14.5}$) \lya forest absorption
systems. Here we present results for four lines-of-sight which contain nearby (${ cz \lesssim 3000\kms}$) \lya absorbers in galaxy
voids. Although our data go quite deep (${ -13 \leqslant M_{B}(limit) \leqslant -14}$) out to impact parameters of ${ \rho \lesssim
100-250\,h_{70}^{-1}\kpc}$, these absorbers remain isolated and thus appear to be truly intergalactic, rather than part of galaxies or
their halos. Since we and others have discovered no galaxies in voids, the only baryons detected in the voids are in the \lya
``clouds''. Using a photoionization model for these clouds, the total baryonic content of the voids is ${ 4.5 \pm 1.5\%}$ of the mean
baryon density.

\end{abstract}


\section{Introduction}

Since the discovery three decades ago of foreground absorption systems in QSO spectra, their origin and nature has been a subject of
ongoing interest, particularly the \lya-only absorption systems.  \citet{bahcall:69a} suggested that the \lya lines arose in the diffuse
gas of extended galaxy halos, while \citet{arons:72a} proposed that they might be primordial material falling into galaxy halos.  More
recently, computer simulations of structure formation \citep{bond:88a,hernquist:96a,miralda:96a,dave:99a} have supported the hypothesis
that these absorbers are not directly related to galaxies but rather are gas associated with a density-varying intergalactic medium
(IGM).

Because the relationship between \lya absorbers and galaxies can only be studied in detail at low-$z$, the { Hubble Space Telescope}
(HST) is required to discover absorbers for this purpose.  The HST QSO Absorption Line Key Project used the Faint Object Spectrograph
(FOS) to detect higher column density absorbers (log ${ N_{HI}\, (\cmsq) \geqslant 14}$) at \mbox{$z = 0$ -} 1.6
\citep{bahcall:93a,weymann:98a}, followed by more sensitive discovery programs with the { Goddard High Resolution Spectrometer} (GHRS)
and { Space Telescope Imaging Spectrograph} (STIS) \citep{morris:91a,impey:99a,tripp:98a} extending to somewhat lower equivalent width
limits (${ \mathcal{W} \geqslant 50 - 100\,{\rm m\AA}}$ at ${z \lesssim 0.3}$).  Finally, in a large (31 sightline) systematic study,
\citet{penton:00a,penton:00b,penton:02a,penton:02b} have used GHRS and STIS at medium resolution (${ \sim 20 \kms}$) to study the lowest
column absorbers (${ 12.3 \leqslant \log N_{HI}\, (\cmsq) \leqslant 14.5}$) at ${ cz \leqslant 30,000\kms}$. These very low-z studies
are particularly useful for understanding the galaxy-absorber relationship, because at such low redshifts even faint dwarf galaxies can
be detected in the vicinity of the absorption systems.

Attempts to find galaxies associated with the higher column density \lya absorbers using the FOS Key Project data \citep[][L95
hereafter]{lanzetta:95a} found that $\sim 30\%$ of the FOS absorbers were associated with bright galaxies. \citet[][C98
hereafter]{chen:98a} extrapolated the L95 results to fainter luminosities to conclude that all ${ \mathcal{W} > 300\,{\rm m\AA}}$
absorbers are associated with galaxies closer than impact parameters of ${\rho = 225\hscale\,{\rm kpc}}$. This extrapolation remains
untested because the bulk of the FOS absorption systems are too far away for fainter galaxies ($L < L^{*}$) to be detected.

Using more sensitive HST/GHRS spectra, Morris et al. (1993) conducted a detailed galaxy survey of the 3C 273 sightline finding mixed
evidence for galaxies near both high and low column density absorbers.  Subsequent studies by \citet{tripp:98a} and \citet{impey:99a}
also found mixed results, even for a subset of 11 absorbers within the Virgo Cluster where the luminosity function of galaxies is
complete to ${ M_B = -16}$ \citep{impey:99a}.

In preliminary studies to the current work, \citet{stocke:95a} and \citet{shull:96a} found low column density \lya absorbers in galaxy
voids.  \citet{stocke:95a} conducted optical and \mbox{H I} ${\rm 21\,cm}$ searches for galaxies near one ``void absorber'', the $cz =
7740\kms$ absorber in the \mbox{Mkn 501} sightline, finding no optical galaxies at $\rho < 106\,\hscale\kpc$ to ${ M_B \geq -16.2}$ and
no \mbox{H I}-emitting galaxies with ${ M_{HI} \geq 6 \times 10^8\,h^{-2}_{70}\,M_\sun}$ within $\rho < 500\hscale\,\kpc$.

Using a sub-sample of our GHRS sightlines, there are 45 low column density \lya absorbers in regions where the CfA redshift survey is
complete to at least $L^*$.  \citet{penton:02a} find that $29 \pm 8\%$ of \lya absorbers lie in galaxy voids, and that, while there is
some evidence that the remaining $\sim$71\% of the absorbers are associated with the same large-scale structure filaments as the
galaxies, there was no definitive association between any \lya absorber and any individual galaxy.  \citet{penton:02a} find no
correlation between impact parameter and equivalent width for ${ \mathcal{W} \lesssim 200\,{\rm m\AA}}$, or for impact parameters
greater than ${ 200\,h_{70}^{-1}\kpc}$, contrary to the predictions of L95 and C98.  However, \citet{penton:02a} do not address the
relationship between absorbers and fainter galaxies.

In order to determine what relationship exists between the low-column \lya absorbers and the faintest dwarf galaxies, we have undertaken
a project based on the absorber sample of \citet{penton:00a, penton:02b}. This sample is ideal because it contains many systems closer
than $cz = 10,000\kms$. At such distances, we are able to probe well down into the galaxy luminosity function with multi-object
spectroscopy on modest aperture telescopes, in some cases reaching the faintest dwarf galaxies (${ M_B \sim -12}$). To achieve this aim
we have undertaken pencil-beam galaxy redshift surveys in the direction of 15 of the 31 QSO sightlines studied with GHRS and STIS by
\citet{penton:00a,penton:02b}.

In this paper we present results from a subset of our data, namely those sightlines in which absorption systems were found inside very
nearby ($cz\leq$ 3000 km s$^{-1}$) galaxy voids, defined by having no bright galaxy (${ M_B \leqslant -17.5}$) closer than
$2\hscale\,{\rm Mpc}$. The study of void absorbers is important because those that lie within bright galaxy filaments are very likely to
have a bright galaxy nearby; fainter galaxies will lie slightly closer, given their higher space density
\citep{impey:99a,penton:02a}. In these cases, it is not clear whether the clouds are associated with individual galaxies or with the
large scale structure in which both the absorbers and galaxies are embedded. Absorbers in galaxy voids are immune to this effect: there
are no bright galaxies nearby, and fainter galaxies are not expected to be close merely by chance.  Thus, these ``void absorbers'' are
uniquely suited to help determine whether all absorbers are galaxy halos, or if some are truly intergalactic material, as is strongly
suggested by simulations ({ e.g.}, Dav\'{e} et al. 1999 \nocite{dave:99a}).

We present our work as follows: In \S 2 we discuss our observational techniques and data reduction. In \S 3 we introduce the void
absorber sightlines and discuss our results as they pertain to the voids. In \S 4 we summarize the conclusions from this work. We assume
${ \hnot = 70\,h_{70} \kms \, Mpc^{-1}}$ throughout.

\section{Observations and Reductions}

These observations were obtained over the past three years at the DuPont ${ 2.5\,{\rm m}}$ telescope of Las Campanas Observatory (LCO)
in Chile and at the ${ 3.5\,{\rm m}}$ WIYN\footnote{The WIYN Observatory is a joint facility of the University of Wisconsin-Madison,
Indiana University, Yale University, and the National Optical Astronomy Observatories.} telescope at Kitt Peak National Observatory
(KPNO) in Arizona. The LCO observations were taken with the Wide Field CCD multi-object spectrometer designed and built by one of us
(R.J.W.). The instrument employs metal slitmasks to obtain redshifts of up to sixty galaxies at a time over a field of view of
approximately $12' \times 20'$. The DuPont spectra range from just shortward of ${ 4000\,{\rm \AA}}$ to ${ 8000\,{\rm \AA}}$, with a
resolution of $\lambda/\Delta \lambda \approx 2000$.  We typically used only the part of the spectrum shortward of ${ 5500\,{\rm \AA}}$
in order to avoid confusion caused by sky subtraction. The WIYN observations were taken with the HYDRA fiber-fed multi-object
spectrometer using the blue fiber cables, the G3 filter and the KPC-007 600 lines/mm grating. The wavelength coverage for these spectra
was from ${ 3800\,{\rm \AA}}$ to ${ 5800\,{\rm \AA}}$ with $\lambda/\Delta \lambda \approx 4000$ at ${ 4000{\rm \AA}}$.  Typically, 60 -
70 spectra were obtained in each fiber setup over a $40'$ circular field of view.  For both LCO and WIYN, we used a set of spectral
templates, kindly provided by E. Ellingson, to measure galaxy redshifts using cross-correlation techniques.  Velocities returned by this
procedure have typical accuracies of ${ \pm 150 \kms}$ for the faintest galaxies with usable redshifts.

Astrometric images were taken either at the ${ 0.9\,{\rm m}}$ telescope of KPNO using the T2kA or MOSAIC detectors or at the Las
Campanas 40 inch telescope. Gunn $r$ or Johnson $R$ images were obtained and galaxy magnitudes converted to Johnson $B$ using the colors
appropriate for an Scd galaxy, as reported in \citet{fukugita:95a}. This is a conservative choice of color conversion since most
galaxies have redder colors and thus fainter $B$ magnitudes for their red magnitudes.  Integration times ensured a depth well below our
desired limiting magnitude of ${ m_r=19}$ (${ m_B \sim 19.7}$).  Object detection and galaxy/star separation were accomplished using
PPP, provided for our use by H. Yee \citep{yee:91a}, or SExtractor \citep{bertin:96a}. Our surface brightness sensitivity varied
somewhat from image to image, but typically we were able to automatically detect galaxies with $\mu_r \geq$ 23 ${\rm mag\,arcsec^{-2}}$
($\mu_B \geqslant 23.7\,{\rm mag\,arcsec^{-2}}$). While these surface brightness limits do not extend to the very lowest values observed
for galaxies in modern surveys, they do extend well into the low surface brightness (LSB) galaxy regime \citep{mcgaugh:95a}. A visual
inspection of the images found a few objects of lower surface brightness which had been missed by the finding algorithm, and which were
added to our spectroscopy lists.  PPP and SExtractor showed adequate agreement in object classifications and magnitudes.
\citet{yee:91a} describes experiments which show that the PPP magnitudes are equivalent to total magnitudes for objects well above the
magnitude limit.

Owing to our desire to obtain spectra of as many galaxies as possible within our survey regions, we were quite conservative with the use
of the object classifiers. We thus obtained spectra of many objects classified as possible galaxies which are, in fact, stars. Since
objects were selected for observation based primarily upon location, so as to maximize the number of objects per mask or fiber setup, we
expect that the fraction of stars among our unobserved objects at each magnitude limit is similar to the fraction among our observed
objects.  Therefore, the galaxy completeness fraction in each field (see Table \ref{tab:voidabs}) takes into account the fraction of
stars mis-classified as possible galaxies in that field.

\section{Results}

We have observed four sightlines from \citet{penton:00a,penton:02b} which contain 7 \lya absorbers in voids (no galaxies at ${ M_B
\leqslant -17.5}$ within ${\rm 2\hscale\,Mpc}$), based upon bright galaxy data from the CfA catalog \citep{huchra:83a}.  Table
\ref{tab:voidabs} gives a summary of our void absorber data, including by column: (1) target name; (2) AGN heliocentric recession
velocity; (3) absorber heliocentric recession velocity; (4) rest equivalent width ($\mathcal{W}$) of the \lya absorption in ${\rm
m\AA}$; (5) apparent $B$ magnitude limit of our spectroscopy; (6) radius of the field of view covered at the distance of the void
absorber; (7) percentage of galaxies for which we have a redshift to ${ m_B(limit)}$ listed in column (5) within the field of view
listed in column (6); and (8) the absolute $B$ magnitude limit of our spectroscopy at the distance of the absorber. The void absorber in
the VII Zw118 sightline is actually a blend of two absorbers with comparable equivalent widths. The combined equivalent width ($333\,
{\rm m\AA}$, as it would appear in an HST/FOS spectrum) of this system places it well within the equivalent width limits suggested by
L95 to be galaxy halos. The Mkn 421 and VII Zw118 fields were observed at WIYN, the other two at LCO. At the bottom of Table
\ref{tab:voidabs} we have added data on the two ``void absorbers'' reported in \citet{stocke:95a} and \citet{penton:02a}.  While farther
away than the other four objects in Table \ref{tab:voidabs}, the results from these two absorbers are entirely consistent with the
results presented herein.

We have detected no galaxy coincident in velocity (${ \pm 300 \kms}$) with an absorber out to substantial impact parameters (${ 100 -
250\hscale\kpc}$, except for HE1029-140, where our galaxy survey work is $\lesssim 10\%$ complete at large impact parameters). In fact,
there are no galaxies detected within the voids containing them. This absence is quite striking: not only have we failed to find
galaxies within several hundred $\kms$ of any absorber, we have found no faint galaxies closer than the nearest bright galaxies that
define the voids. Thus, even if we use the \lya absorbers as markers for the presence of material in voids, a sensitive search for
galaxies near these absorbers fails to find even very faint galaxies. These observations underscore previous negative searches for
galaxies in voids \citep{popescu:97a,szomoru:96b, szomoru:96a}.

\begin{figure}
\epsscale{0.6}\plotone{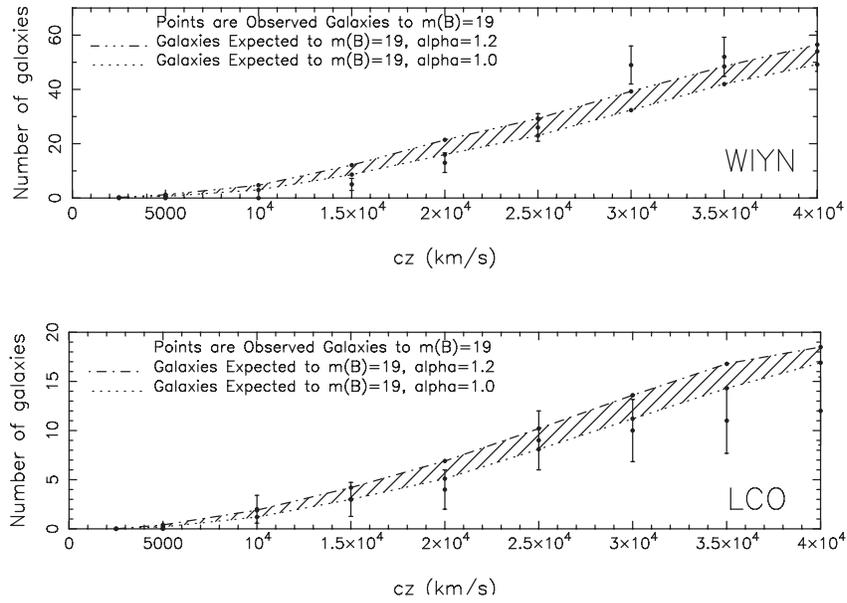}
\caption{The cumulative number of
galaxies expected compared to the number observed as a function of velocity 
($cz$) for our survey. We have used the functional form and
normalizations from Schechter (1976) with different faint-end slope, $\alpha$, 
to compute the expected number of galaxies. The expected
values are plotted in dashed-dots on the upper bound of the shaded region, and 
dots on the lower bound, for \mbox{$\alpha=-1.2$} and
$-1.0$, respectively. Both curves assume a limiting magnitude of $m_B$ = 19. 
The top plot is for two WIYN fields, and the bottom for
five fields taken at Las Campanas. The points and error bars mark the number 
of galaxies we have found in our sample, along with
$\sqrt{n}$ uncertainties.}
\label{fig:ngal}
\end{figure}

We have tested the sensitivity of our observational procedure by comparing the number of galaxies detected to that expected in 7 of our
survey sightlines (five are from LCO and two are from WIYN) which are $\gtrsim 80\%$ complete to ${ m_B \approx 19}$ out to impact
parameters $\rho \gtrsim 5'$. We assume a standard Schechter (1976) luminosity function and normalizations for this comparison, with
faint-end slope of either $\alpha = - 1.0$ or $- 1.2$.  Out to $cz = 40,000 \kms$, the number of galaxies detected by our survey agrees
well with the number predicted (see Figure \ref{fig:ngal}). Our success in measuring redshifts for faint galaxies gives us confidence
that we are not missing objects simply because our program lacks the required sensitivity.

While we cannot rule out the possibility that very LSB galaxies ($ \mu_B \geqslant 24\,{\rm mag\,arcsec^{-2}}$) might be responsible for
the void absorbers \citep{linder:98a}, our surface brightness limits are low enough to require that any undetected galaxies in these
regions are extreme LSBs \citep{mcgaugh:95a}. \citet{rauch:96a} did not find any LSB galaxy associated with two very nearby absorbers in
the 3C 273 sightline, nor did \citet{impey:99a} find any LSB galaxies in the Virgo Cluster associated with absorbers.  Recent
observations by \citet{bowen:01a} find \lya absorption in the PKS1004+130 sightline $200\hscale\kpc$ from an LSB galaxy.  These authors
find a high surface brightness galaxy closer to the sightline and interpret the complex \lya absorption as arising in an intra-group
medium, not from a combination of galaxy halos.  \mbox{H I} observations could detect gas-rich LSB galaxies that might be missed in
optical studies \citep{szomoru:96a,szomoru:96b,spitzak:98a,zwaan:97a, rosenberg:00a}.  Van Gorkom et al. (1993, 1996)
\nocite{vangorkom:96a,vangorkom:93a} used Westerbork and the VLA to search for \mbox{H I} toward nine \lya absorbers in four sightlines,
detecting galaxies close to four of them. However, in only one case, that of a dwarf (${ M_B = -16}$) toward Mkn 335, is the projected
distance from the sightline to the galaxy ${ \lesssim 100\,h_{70}^{-1}\kpc}$.  But this object has a high central surface brightness
($m_B \approx 19.5\,{\rm mag\,sec^{-2}}$) and would not have been missed by this survey. For the other five absorbers, the nearest
galaxy neighbors are several hundred kpc away. At these distances other galaxies become comparably close, making association with a
single galaxy halo problematic \citep{shull:96a,shull:98a,penton:02a}.

Van Gorkom et al. (1996) \nocite{vangorkom:96a} find that the rate of \mbox{H I} detections is not affected by the presence of \lya
absorbers, evidence that the matches they have found between absorbers and galaxies are likely to be chance associations rather than
actual individual physical connections. New VLA \mbox{H I} observations have targeted three of the void absorbers in Table
\ref{tab:voidabs}, with sufficient sensitivity to address the \mbox{H I}-rich LSB galaxy possibility definitively \citep{hibbard:02a}.
Thus, while we cannot exclude categorically the presence of extreme LSB galaxies near these void absorbers, there is no positive
evidence for association between any \lya forest absorber and an LSB galaxy to date.

\section{Summary and Conclusions}

We have presented a subset of the results from our pencil-beam redshift survey toward 15 AGN containing nearby \lya absorbers: namely,
four sightlines that pass through very nearby (${cz \leqslant 3000 \kms}$) galaxy voids.  The absorbers in these voids are all
sufficiently near that we could have detected faint dwarf galaxies to limits of ${ M_B \lesssim -14.3}$ or fainter within ${
100-250\kpc}$ of the sightline. In one sightline, HE1029-140, our survey is too incomplete to be definitive.  The remaining fields are
more than 90\% complete for ${ M_B \leqslant -12.8}$ to $-14.5$, or to $\mu_B \geqslant 23.5\,{\rm mag\,arcsec^{-2}}$. While our survey
puts strong constraints on the optical luminosity of any normal surface brightness galaxy that could be associated with the void
absorbers, the possibility remains that extreme LSB galaxies ($\mu_B \geqslant 24\,{\rm mag\,arcsec^{-2}}$) could be present. A 21\,cm
search has been made for \mbox{H I} near three void absorbers \citep{hibbard:02a} and will address the presence of hydrogen-rich LSBs
near the void absorbers.

The results of this optical survey strongly suggest that the void absorbers are a primordial population of objects. This is in keeping
with results from cosmic structure formation simulations (e.g., Dav\'{e} et al. 1999\nocite{dave:99a}) and is contrary to the hypothesis
of C98 that all \lya absorbers are associated with galaxies. The primordial nature of these void absorbers can be definitively addressed
by searching for metals in them.

The current observations allow a first determination of the total baryon content in voids. Since no galaxies have been found by us or
others in voids, the only baryons detected inside the voids are the \lya absorbing clouds. Based upon a standard photoionization model,
\citet{penton:02a} have determined that the local \lya forest accounts for $\sim 20\%$ of the total baryons predicted by big bang
nucleosynthesis. By comparing their measured value of $dN/dz$ outside of voids to that in the voids, and by assuming the same values for
the local ionizing radiation flux and the sizes of \lya clouds in both void and non-void absorbers, \citet{penton:02b} show that the
void absorbers account for $\sim 22\pm 8\,\%$ of local absorption per unit pathlength (i.e., the $29\%$ quoted earlier, corrected for
the relative amount of pathlength through voids in the survey). This implies that ${ 4.5\pm 1.5\,\%}$ of the total baryons reside in the
voids. This calculation assumes that all baryons in voids are detectable either as galaxies or \lya absorbers; i.e., if a warm-hot IGM
component is present without detectable \lya absorption, we have not accounted for it.

\acknowledgments

Support at Colorado for work on the local \lya forest comes from grants GO-06593.01 and GO-08182.01 from the Hubble Space Telescope
Science Institute.

\newcommand{\SortNoop}[1]{}

\small
\begin{deluxetable}{lcccrccc}
\tabletypesize{\small}
\tablecolumns{8}
\tablewidth{0pc} 
\tablecaption{Galaxy Void Absorbers} 
\tablehead{ 
\colhead{AGN}               &
\colhead{${cz_{AGN} }$ }    & 
\colhead{${cz_{abs} }$ }    & 
\colhead{$\mathcal{W}$}     & 
\colhead{${m_B}$}           & 
\colhead{Radius }           & 
\colhead{Complete}          &
\colhead{${M_B}$} \\
 				& (\kms) & (\kms)&   (m{\rm \AA})     &  (limit)  	&(kpc)	&   (\%)   &  
(limit)  }
\startdata 
Mkn 421         		& 9000   & 3035  & 87$\pm$15   	& 18.7 & 252 	& 95   	& 
-14.5     \\
Mkn 509         		& 10,312 & 2548  & 211$\pm$32  	& 20.0 & 105 	& 96  	& -12.8 
    \\
VII Zw 118      		& 23,881 & 2426  & 189$\pm$151 	& 18.6 & 201 	& 91   	& 
-14.0     \\
VII Zw 118      		& 23,881 & 2469  & 144$\pm$116	& 18.6 & 205 	& 91   	& -14.0 
    \\
HE 1029-140     		& 25,782 & 1979  & 103$\pm$45  	& 18.8 & 41  	& 70  	& -13.3 
    \\
\hline
Mkn 501\tablenotemark{1}        & 10,092 & 5990  & 55$\pm$46    & 19.0 & 82  	
& 100  	& -15.7     \\
                                &        & 7740  & 53$\pm$36    & 19.0 & 106 	
& 100  	& -16.2     \\
\enddata\label{tab:voidabs}
\tablenotetext{1}{data from \citet{stocke:95a} }
\end{deluxetable}

\end{document}